\documentclass[aps,twocolumn,prl]{revtex4}
\usepackage{graphicx,amssymb,amsmath}
\usepackage{graphicx}
\usepackage{dcolumn}
\usepackage{bm}

\begin{document}
\title{Precursors to Exciton Condensation in Quantum Hall Bilayers}
\author{J. P. Eisenstein}
\affiliation{Institute for Quantum Information and Matter, Department of Physics, California Institute of Technology, Pasadena, CA 91125, USA}
\author{L. N. Pfeiffer}
\affiliation{Department of Electrical Engineering, Princeton University, Princeton, NJ 08544, USA}
\author{K. W. West}
\affiliation{Department of Electrical Engineering, Princeton University, Princeton, NJ 08544, USA}

\date{\today}

\begin{abstract}
Tunneling spectroscopy reveals evidence for interlayer electron-hole correlations in quantum Hall bilayer two-dimensional electron systems at layer separations near, but above, the transition to the incompressible exciton condensate at total Landau level filling $\nu_T=1$.  These correlations are manifested by a nonlinear suppression of the Coulomb pseudogap which inhibits low energy interlayer tunneling in weakly-coupled bilayers.  The pseudogap suppression is strongest at $\nu_T=1$ and grows rapidly as the critical layer separation for exciton condensation is approached from above.  
\end{abstract}

\maketitle
Theoretical suggestions\cite{blatt1962,moskalenko1962,keldysh1965,jerome1967} for Bose condensation of excitons first emerged in the decade following the Bardeen-Cooper-Schrieffer theory\cite{bardeen1957} of superconductivity.  Almost four decades elapsed before strong experimental evidence for such condensation began to accumulate.   Initially this evidence came from tunneling and transport experiments on bilayer two-dimensional electron systems in which stable exciton populations emerge at high magnetic field\cite{spielman2000,kellogg2002} and from photo-luminescence experiments on transient exciton populations in coupled quantum wells\cite{butov2002,snoke2002}.  Recently\cite{kogar2017}, exciton condensation has been detected via electron energy loss spectroscopy on a three-dimensional solid, the transition metal dichalcogenide semimetal 1{\it T}-TiSe$_2$.

In the bilayer two-dimensional electron system (2DES) case, the exciton condensate appears when the total number of electrons matches the number of available states in a single spin-resolved Landau level created by the magnetic field.  In the density balanced case, each layer contains a 2DES at half filling of the lowest Landau level (LL).  If the layers are sufficiently close together and the temperature is sufficiently low, interlayer Coulomb interactions stabilize a remarkable broken symmetry phase in which electrons are shared equally between the two layers, even in the hypothetical absence of zero interlayer single particle tunneling.  In addition to a quantized Hall plateau at $\rho_{xy}=h/e^2$, this phase displays several other fascinating properties, including Josephson-like interlayer tunneling, quantized Hall drag, and nearly dissipationless transport of counterpropagating currents across the bulk of the 2D system \cite{eisenstein2004}. There are multiple equivalent ways to describe this phase, including as an easy-plane ferromagnet or as a condensate of interlayer excitons. Of course, interactions between electrons within the same layer are strong independent of the layer separation and, in the large separation limit, each 2DES at half filling of the lowest LL is a compressible, nonquantized Hall phase well described as a Fermi liquid of composite fermions \cite{halperin1993,jain1989}.  As the layer separation $d$ is reduced, interlayer Coulomb interactions become increasingly important and this description breaks down.  At some critical layer separation $d_c$ a transition to the incompressible exciton condensate occurs.  The nature of this transition, and of the bilayer 2DES generally at $d \gtrsim d_c$, remain poorly understood despite intensive and ongoing study\cite{cote1992,bonesteel1996,kim2001,veillette2002,simon2003,doretto2006,shibata2006,park2006,moller2008,alicea2009,cipri2014,zhu2017,sodemann2017,lian2018}.

Here we report evidence from interlayer tunneling spectroscopy experiments that significant interlayer particle-hole correlations exist in bilayer 2D electron systems at layer separations larger than those required for exciton condensation. These correlations are strongest when the per-layer LL filling fraction is $\nu = 1/2$ and grow in importance as the effective layer separation is reduced and the excitonic transition approached.  In this regime the bilayer 2DES is compressible, exhibits no quantized Hall plateau and neither ordinary longitudinal nor Hall drag transport presents any significant anomaly.  In contrast, interlayer tunneling is well-suited to exploring subtle interlayer particle-hole correlations in part because in their absence the tunneling rate is heavily suppressed by intralayer Coulomb interactions\cite{hatsugai93,he93,johannson93,haussmann96,levitov97,chowdhury2018}.

\begin{figure}
\includegraphics[width=3.2in]{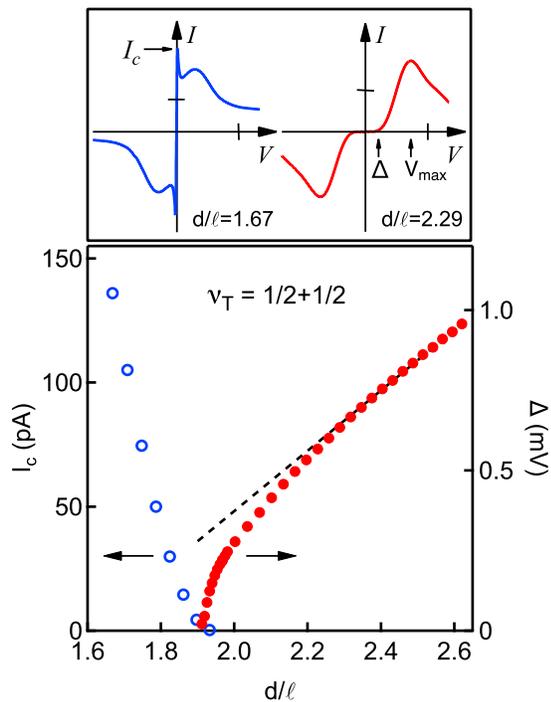}
\caption{Aspects of interlayer tunneling at $\nu_T=1/2+1/2$ at $T=50$ mK.  Upper panel: Typical tunneling $IV$ curves at effective layer separations $d/\ell$ above (right) and below (left) the transition to the excitonic phase.  (The voltage and current tick marks are at 3 mV and 50 pA, respectively.) Lower panel:  Red dots: Collapse of pseudogap $\Delta$ as $d/\ell$ is reduced.  Blue open dots: Tunneling critical current in excitonic phase.}
\end{figure}

Figure 1 shows two interlayer tunneling current-voltage ($IV$) characteristics observed in a single, density balanced, bilayer 2DES sample containing two 18 nm GaAs quantum wells separated by a 10 nm AlGaAs barrier layer.   (We here discuss only tunneling between the lowest LLs in each layer.)  For the left trace the 2DES density $n$ in each layer has been electrostatically tuned to be relatively low, while for the right trace it is relatively large.  In each case a perpendicular magnetic field $B_\perp$ yielding $\nu=nh/eB_\perp =1/2$ in each 2D layer has been applied.  Owing to the different densities and magnetic fields the effective layer separation $d/\ell$ (with $d=28$ nm the center-to-center quantum well separation and the magnetic length $\ell=(\hbar/eB_\perp)^{1/2}$) is $d/\ell=1.67$ for the left trace and 2.29 for the right trace.  The left $IV$ characteristic displays the Josephson-like jump in the tunneling current at $V=0$ associated with the quantum Hall exciton condensate\cite{spielman2000,tiemann08}, while the right $IV$ curve shows a pronounced suppression\cite{eisenstein92,brown94,ashoori90} of the current around $V=0$.  While this suppression can be qualitatively understood as a pseudogap arising from the inability of a strongly correlated single layer 2DES to rapidly relax the charge defects created by the near-instantaneous injection (or extraction) of a tunneling electron at high magnetic field \cite{hatsugai93,he93,johannson93,haussmann96,levitov97,chowdhury2018}, it is our purpose here to demonstrate that interlayer particle-hole correlations modify this picture significantly. 

The transition between the two types of $IV$ chararacteristics at total filling factor $\nu_T=1/2+1/2=1$ is quantitatively illustrated in the lower panel of Fig. 1.  The red solid dots show the dependence of the voltage width $\Delta$ of the suppressed region of tunneling around $V=0$ on the effective layer separation $d/\ell$.  (We define $\Delta$ as the voltage where the tunneling current rises to 2\% of the maximum current observed at $V=V_{max}$.)  The blue open dots show the magnitude $I_c$ of the Josephson-like current jump at $V=0$ observed in the excitonic phase.  The figure demonstrates that the collapse of the tunneling pseudogap $\Delta$ and onset of Josephson-like interlayer tunneling occur at essentially the same effective layer separation, about $d/\ell\approx 1.93$ in the present sample.  

The dashed straight line in the lower panel of Fig.1 emphasizes the increasing nonlinearity of the $\Delta$ vs. $d/\ell$ dependence as the excitonic transition is approached.  Since $\ell^{-1}=(2\pi n/\nu)^{1/2}$, $\Delta$ is similarly nonlinear in $n^{1/2}$.  This is perhaps surprising since in the simplest scenario lowest LL tunneling between widely separated 2D layers is dominated by intralayer Coulomb interactions\cite{hatsugai93,he93,johannson93,haussmann96,levitov97,chowdhury2018} which scale linearly with $n^{1/2}$ at fixed $\nu$.

Figure 2 contrasts this unusual nonlinear dependence of $\Delta$ upon $n^{1/2}$ at $\nu_T=1/2+1/2$ with the $linear$ dependence more commonly observed.  Figure 2(a) presents the $n^{1/2}$ dependence of $V_{max}$, the voltage location of the peak tunnel current.  The red solid dots are from the same sample, and at the same densities, as the $\Delta$ data shown in Fig. 1, while the open dots are from a second sample in which the width of the tunnel barrier has been increased from $d_b=10$ to 38 nm (thus doubling $d$, the center-to-center quantum well separation, from 28 to 56 nm.)
In both samples $V_{max}$ exhibits a clear linear dependence on $n^{1/2}$ which extrapolates to a negative intercept in the $n \rightarrow 0$ limit.  (This negative intercept reflects the attraction, \emph{in the final state}, between a tunneled electron and the hole it leaves behind in the source layer.  The attraction is of course weaker in the wider barrier sample and this accounts for the roughly vertical displacement of the two data sets \cite{comment2}.  This final state effect is not to be confused with interlayer electron-hole correlations present in the \emph{initial} state of the bilayer 2DES.)

\begin{figure}
\includegraphics[width=3.2in]{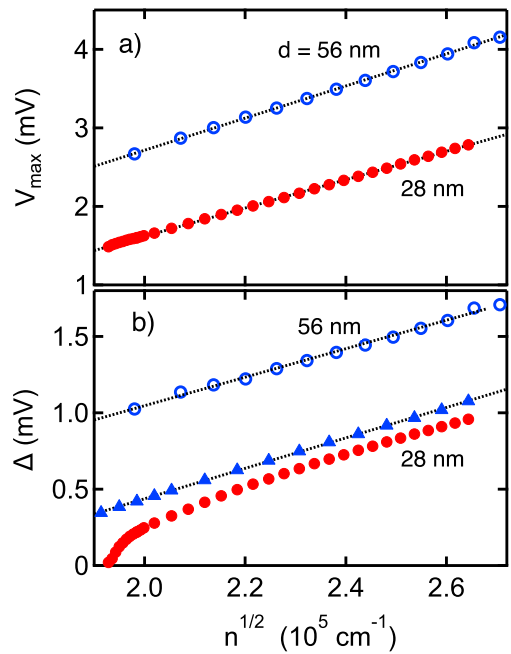}
\caption{Density dependences of $V_{max}$, the voltage at which the tunnel current is maximized, and the pseudogap $\Delta$, in two samples having different layer separations $d$.  (a) $V_{max}$ at $\nu_T=1/2+1/2$.  (b) $\Delta$ at $\nu_T=1/2+1/2$ (open blue and red solid dots) and at $\nu_T=0.41+0.41$ (triangles).  All data are at $T=50$ mK.}
\end{figure}

Figure 2(b) returns to the pseudogap $\Delta$, as defined above.  The open dots are the $\Delta$ values, obtained at $\nu_T=1/2+1/2$, from the wide barrier sample.  The density range is the same as for the $\Delta$ values obtained from the narrow barrier sample shown in Fig. 1 and repeated in Fig. 2(b) (red solid dots) for ease of comparison.  Unlike the nonlinear collapse of $\Delta$ seen in the $d=28$ nm  sample, $\Delta$ in the $d=56$ nm sample exhibits a simple linear dependence on $n^{1/2}$.  These very different dependences strongly suggest that interlayer Coulomb interactions, which eventually lead to exciton condensation in the narrow barrier sample but not in the wide barrier sample, are, especially at low density, strong in the former but weak in the latter \cite{comment8}.  That at the highest densities the slopes $d\Delta/d(n^{1/2})$ become roughly equal is not surprising since intralayer interactions then dominate over interlayer interactions.

Finally, the solid triangles in Fig. 2(b) are the $\Delta$ values, in the narrow barrier sample, obtained when each two-dimensional layer is at filling factor $\nu_T = 0.414+0.414 = 0.828$.  This total filling factor is well removed from $\nu_T=1$ where exciton condensation is observed, and is midway between $\nu_T=2/5+2/5$ and $\nu_T=3/7+3/7$ where fractional quantum Hall states exist \cite{comment3}.  As the figure shows, we find $\Delta$ to be linear in $n^{1/2}$ at this filling factor.

\begin{figure}
\includegraphics[width=3.3in]{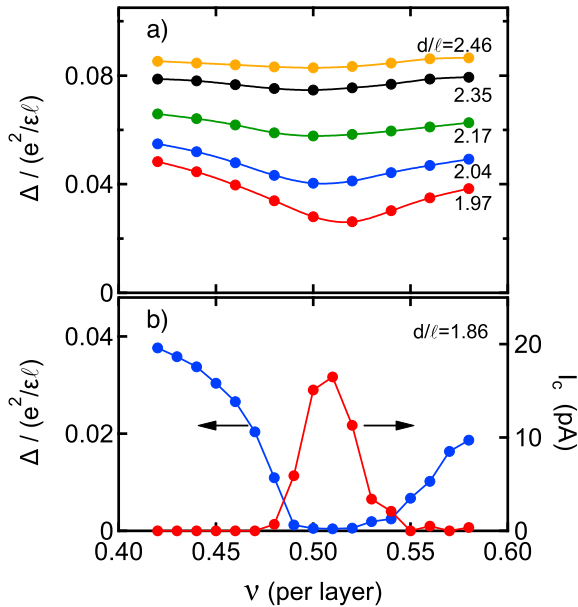}
\caption{(a) Tunneling pseudogap $\Delta$ vs filling factor $\nu$ (same in both layers) at various layer densities $n$ in the $d=28$ nm sample.  $\Delta$ is normalized by the Coulomb energy $e^2/\epsilon\ell$ at each $\nu$.  The data sets are labeled by the effective layer separation $d/\ell$, computed at $\nu=1/2$. (b) At still lower density, the $\Delta$ collapses to zero and a Josephson-like current jump $I_c$ emerges in a window around $\nu = 1/2$. } 
\end{figure}

Further evidence that the nonlinear dependence of $\Delta$ on $n^{1/2}$ in the narrow barrier sample is keyed to total filling factor $\nu_T=1$ is presented in Fig. 3(a).  Here $\Delta$, normalized by the Coulomb energy $e^2/\epsilon \ell$, is plotted versus the per-layer filling factor $\nu$ (the same in both layers).  The various traces, which correspond to different 2DES densities $n$, are labeled by the $d/\ell$ value at $\nu_T=1/2+1/2$.  At all densities $\Delta$ vs $\nu$ exhibits a local minimum close to $\nu=1/2$.  The minimum is weak, but clearly observable, at $d/\ell=2.46$ but rapidly deepens as the density is reduced toward $(d/\ell)_c \approx 1.93$ where, at $\nu_T=1$, exciton condensation and the first indications of a Josephson-like zero bias tunneling anomaly appear.  Indeed, as Fig. 3(b) shows, at $d/\ell = 1.86$ $\Delta$ collapses to zero and a Josephson-like zero bias current jump emerges around $\nu = 1/2$.

These data demonstrate that in spite of the generally strong suppression of low energy tunneling between parallel two-dimensional electron systems at high magnetic field, at $\nu_T=1/2+1/2$ this suppression can itself be suppressed, and low energy electrons tunnel more freely, if the separation between the layers is not too large.  This effect is detectable at fairly large layer separation, $d/\ell \sim 2.5$, where the bilayer 2DES is in a compressible, nonquantized Hall state, and becomes stronger as $d/\ell$ is reduced.

Interlayer electron-hole correlations suggest at least a partial explanation for our observations\cite{comment4}.  If electrons in either layer are always accompanied by a strong correlation hole in the opposite layer, the resulting interlayer dipolar electric field presumably lowers the effective tunnel barrier.  Moreover, the strength of the correlation hole undoubtedly grows as the layer separation is reduced.  While such a correlation hole presumably exists at essentially all compressible filling factors, $\nu_T=1/2+1/2$ is special insofar as even in the absence of Coulomb interactions there is an equal number of  unoccupied lowest LL orbitals in one layer and occupied orbitals in the other.  The above model, however, does not readily account for the clear indications in Figs. 1 and 3 that the collapse of the pseudogap $\Delta$ is related to the emergence of the $\nu_T=1$ exciton condensate.  Indeed, the gap $\Delta$ collapses to zero at essentially the same $d/\ell$ as where the first signs of Josephson-like tunneling (and other signature phenomena, such as quantized Hall drag) appear.  
This behavior suggests that the nonlinear collapse of $\Delta$ reflects excitonic fluctuations in anticipation of exciton condensation at lower layer separations.

We turn now to the effect of layer density imbalance on the tunneling $IV$ characteristic.  Via electrostatic gating the filling factors $\nu_1$ and $\nu_2$ of the individual two-dimensional layers can be adjusted so that $\nu_T = \nu_1+\nu_2=1$ but $\Delta \nu \equiv \nu_1-\nu_2 \neq 0$. Not surprisingly, nonzero $\Delta \nu$ alters the tunneling $IV$ curve.  In the absence of significant interlayer correlations, a simple, if crude, model of the tunneling pseudogap illustrates this:  For an electron to tunnel from layer 1 to layer 2 and overcome the pseudogap, the interlayer voltage must be at least as large as $e|V_{1,2}| \sim \epsilon^-(\nu_1)+\epsilon^+(\nu_2)$, where $\epsilon^-$ and $\epsilon^+$ are the energies required to rapidly extract and inject an electron into a strongly correlated 2DES. Similarly, in the opposite bias polarity, the minimum voltage required for tunneling from layer 2 to layer 1 would be $e|V_{2,1}| \sim \epsilon^-(\nu_2)+\epsilon^+(\nu_1)$.  Since $\epsilon^-(\nu)$ and $\epsilon^+(\nu)$ are in general different \cite{haussmann96}, these voltage thresholds are also different, unless $\nu_1=\nu_2$.

Figure 4 displays the pseudogaps $\Delta_+$ and $\Delta_-$, determined separately from the positive (red dots) and negative (black open dots) voltage portions of the $IV$ curve, versus $\Delta \nu = \nu_1 - \nu_2$ at $d/\ell=2.46$ and $d/\ell=2.00$  \cite{comment6,comment7}.  As expected, at both $d/\ell$ values, $\Delta_+$ and $\Delta_-$ are closely equal at $\Delta \nu=0$ where the bilayer is density balanced.  However, at finite density imbalance the pseudogap behaves very differently at high and low $d/\ell$.  At $d/\ell=2.46$, where Fig. 3(a) suggests that interlayer electron-hole correlations are present but weak, $\Delta_+$ and $\Delta_{-}$ separate from one another roughly linearly with $\Delta \nu$.  This is consistent with the crude model of tunneling between independent layers described above.  In contrast, at $d/\ell=2.00$ the pseudogaps $\Delta_+$ and $\Delta_-$ remain nearly equal and {\it decrease}, roughly as $|\Delta \nu|^2$, as the bilayer is imbalanced.  Although this imbalance-induced reduction of the pseudogap is not well understood, it is again likely related to proximity to the $\nu_T=1$ exciton condensate.  Indeed, experiments \cite{spielman04,wiersma04,clarke05,champagne08} have shown that the critical layer separation for exciton condensation $increases$ slightly with density imbalance.  Hence, in analogy to the non-linear collapse of $\Delta$ near $d/\ell \approx 1.93$ observed in density balanced $\nu_T=1$ bilayers (shown in Fig. 1), a small density imbalance would likely yield a similar collapse, only shifted to slightly larger $d/\ell$.  In that case, at a fixed $d/\ell$ near, but above, the collapse point, $\Delta$ at imbalance $\Delta \nu \neq 0$ would be smaller than in the density balanced $\Delta \nu = 0$ case.  This is consistent with the data shown in Fig. 4(b).   We emphasize that while $d/\ell=2.00$ is close to the critical layer separation, the bilayer remains in the incoherent $\nu_T=1$ phase at all $\Delta \nu$ examined; i.e. no Josephson-like tunneling anomaly is observed. 

\begin{figure}
\includegraphics[width=3.2 in]{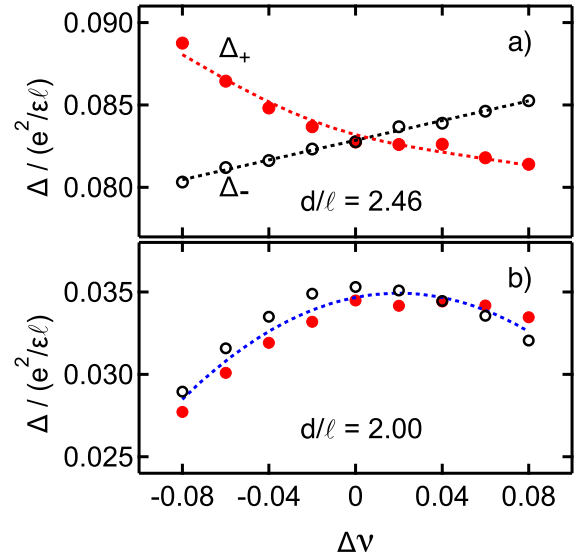}
\caption{Tunneling pseudogap $\Delta_{+,-}$ at $\nu_T=1$ measured at positive (red dots) and negative (black open dots) interlayer voltage vs layer density imbalance $\Delta \nu \equiv \nu_1-\nu_2$ at $d/\ell=2.46$ and 2.00.  The dashed lines in (a) are guides to the eye; in (b) it is a parabolic fit to the average of $\Delta_+$ and $\Delta_-$.  At positive (negative) interlayer voltage electrons tunnel from layer 2 (1) to layer 1 (2).}
\end{figure}

While the above results are suggestive of a second order phase transition, there is also evidence that the transition may be first order\cite{schliemann01,spielman05,kumada05}.  For example, experiments\cite{spielman05,kumada05} have demonstrated that the critical layer separation for exciton condensation increases slightly when the electronic spin Zeeman energy is enhanced via the hyperfine coupling to the nuclear spins of the host lattice. Zou {\it et al.}\cite{zou10} found that this is at least consistent with a first order phase transition in which the spin polarization of the bilayer 2DES jumps discontinuously at the critical point.  

In conclusion, the various tunneling data presented here suggest the presence of interlayer electron-hole correlations at layer separations significantly larger than that required for observation of the key features of the $\nu_T=1$ exciton condensate.  These correlations, which manifest as a suppression of the tunneling pseudogap, are strongest at $\nu_T=1$ and gather in strength as the excitonic phase is approached.  Moreover, their dependence on layer density imbalance is consistent with the known imbalance dependence of the excitonic phase boundary.  These observations point to fluctuations of the excitonic phase persisting into the compressible phase well above the critical layer separation.  

We thank G. Chaudhary, S. Das Sarma, P.A. Lee, A.H. MacDonald, and G. Refael for helpful discussions.
This work was supported in part by the Institute for Quantum Information and Matter, an NSF Physics Frontiers Center with support of the Gordon and Betty Moore Foundation through Grant No. GBMF1250.  The work at Princeton University was funded by the Gordon and Betty Moore Foundation through Grant GBMF 4420, and by the National Science Foundation MRSEC Grant 1420541.


\begin{references}
\bibitem{comment8}  In principle, exciton condensation might occur in the wide barrier sample as well, but at a density four times lower than in the narrow barrier sample.  Disorder would likely defeat this possibility in our present samples.
\bibitem{comment3} These fractional quantum Hall states are very weak in the present samples.
\bibitem{comment4} The role of interlayer correlations in the tunneling between 2DESs has been considered theoretically previously\cite{efros93,varma94}.
\bibitem{comment6}  The imbalance $\Delta \nu$ is here estimated from density calibrations done at very low magnetic field.  This ignores
the small many-body corrections arising at high magnetic field discussed in Ref. \cite{champagne08}.
\bibitem{comment7} The tunneling set-up is such that for $V>0$ electrons tunnel from layer 2 to layer 1.
\end{references}
\end{document}